\documentclass{article}

\usepackage{microtype}
\usepackage{graphicx}
\usepackage{booktabs} 

\usepackage{hyperref}
\usepackage{enumitem}

\usepackage[accepted]{icml2024}

\usepackage{amsmath}
\usepackage{amssymb}
\usepackage{mathtools}
\usepackage{amsthm}

\usepackage[capitalize,noabbrev]{cleveref}

\usepackage{url}
\usepackage[frozencache,cachedir=mintedcache]{minted}
\usepackage{inconsolata}
\usepackage{xspace}
\usepackage{nolbreaks}
\usepackage{ulem}
\usepackage{booktabs}
\usepackage{subcaption}

\theoremstyle{plain}

\theoremstyle{definition}

\theoremstyle{remark}

\usepackage[textsize=tiny]{todonotes}

\icmltitlerunning{AST-T5: Structure-Aware Pretraining for Code Generation and Understanding}

\newcommand{\model}{\nolbreaks{AST-T5}\xspace}

\begin{document}

\twocolumn[
\icmltitle{AST-T5: Structure-Aware Pretraining for Code Generation and Understanding}



\icmlsetsymbol{equal}{*}

\begin{icmlauthorlist}
\icmlauthor{Linyuan Gong}{ucb}
\icmlauthor{Mostafa Elhoushi}{meta}
\icmlauthor{Alvin Cheung}{ucb}
\end{icmlauthorlist}

\icmlaffiliation{ucb}{University of California at Berkeley}
\icmlaffiliation{meta}{AI at Meta}

\icmlcorrespondingauthor{Linyuan Gong}{gly@berkeley.edu}

\icmlkeywords{PLACEHOLDER: Keywords}

\vskip 0.3in
]



\printAffiliationsAndNotice{}  

\begin{abstract}
Large language models (LLMs) have made significant advancements in code-related tasks, yet many LLMs treat code as simple sequences, neglecting its structured nature. We introduce \model, a novel pretraining paradigm that leverages the Abstract Syntax Tree (AST) for enhanced code generation, transpilation, and understanding. Using dynamic programming, our AST-Aware Segmentation retains code structure, while our AST-Aware Span Corruption objective equips the model to reconstruct various code structures. Unlike other models, \model avoids complex program analyses or architectural changes, so it integrates seamlessly with any encoder-decoder Transformer. Evaluations show that \model consistently outperforms similar-sized LMs across various code-related tasks including HumanEval and MBPP. Structure-awareness makes \model particularly powerful in code-to-code tasks, surpassing CodeT5 by 2 points in exact match score for the Bugs2Fix task and by 3 points in exact match score for Java-C\# Transpilation in CodeXGLUE. Our code and model are publicly available at
\ifdefined\isaccepted
\url{https://github.com/gonglinyuan/ast\_t5}
\else
\url{https://annonymized}
\fi.
\end{abstract}

\section{Introduction}\label{sec:introduction}

\begin{figure*}[ht]
\vskip 0.2in
\begin{center}
\centerline{\includegraphics[]{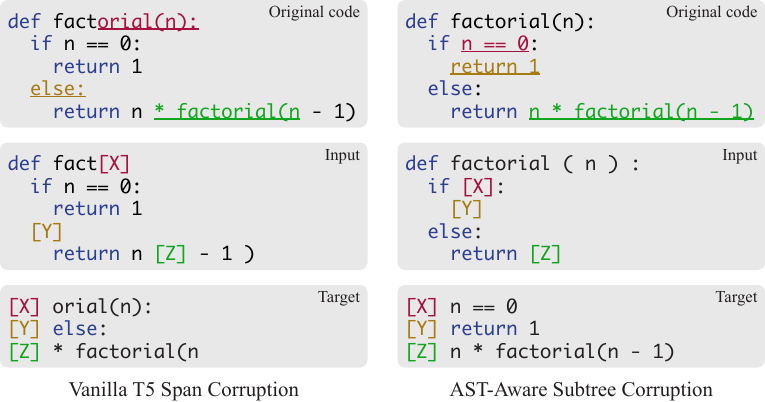}}
\caption{Comparison of AST-Aware Subtree Corruption and Vanilla T5 using a Python factorial function. Both methods replace masked spans with sentinel tokens (special tokens added to the vocabulary, shown as \textcolor[HTML]{A61140}{\texttt{[X]}}, \textcolor[HTML]{A67B11}{\texttt{[Y]}}, and \textcolor[HTML]{11A61D}{\texttt{[Z]}} in the figure), with output sequences containing the original masked tokens. Inputs and targets are shown in byte-pair encoding (BPE); for instance, ``factorial'' is encoded into ``fact'' and ``orial''. Unlike Vanilla T5, which masks random spans without considering code structure, our approach specifically targets spans aligned with AST subtrees, like expressions and statements.}\label{fig:ast_aware_span_corruption}
\end{center}
\vskip -0.2in
\end{figure*}

We have witnessed the transformative impact of large language models (LLMs) on various aspects of artificial intelligence in recent years~\citep{gpt3,instructgpt,llama}, especially in code generation and understanding~\citep{codebert,codet5,codellama}. By pretraining on massive code corpora such as the GitHub corpus, LLMs learns rich representations, thereby becoming powerful tools for various downstream applications such as text-to-code generation~\citep{codex,mbpp,concode}, code-to-code transpilation~\citep{codexglue,transcoder,bugs2fix}, and code understanding (mapping code to classification labels)~\citep{devign,bigclonebench}.

Despite these impressive advances, most existing models interpret code as mere sequences of subword tokens, overlooking its intrinsic structured nature. Prior research has shown that leveraging the Abstract Syntax Tree (AST) of code can significantly improve performance on code-related tasks~\citep{graphcodebert,structcoder}. Some studies also use code obfuscation during pretraining to teach models about abstract code structures~\citep{dobf,codet5}. However, these models often rely on computationally expensive processes like Control-Flow Analysis (CFA), obfuscation, or even actual code execution. Such dependency limits their scalability and imposes stringent conditions like code executability. Consequently, these methods may struggle with real-world code, especially in intricate languages like C/C++, where comprehensive analysis remains elusive.

In this study, we propose \model, a pretraining paradigm that leverages the Abstract Syntax Tree (AST) structure of code. The key contribution in \model is a simple yet effective way to exploit code semantics, without the need to run expensive program analysis or execution. Using a lightweight, multi-language parser called Tree-sitter\footnote{\url{https://tree-sitter.github.io/tree-sitter/}}, our approach has broad applicability across all syntactically well-defined programming languages. After we parse code into ASTs, we use a dynamic programming-based segmentation algorithm for AST-aware code segmentation to maintain the structural integrity of the input code. Using our novel AST-Aware Span Corruption technique, the model is pretrained to reconstruct various code structures, ranging from individual tokens to entire function bodies. Together, our approach offers three key advantages: (1) enriched bidirectional encoding for improved code understanding, (2) the ability to coherently generate code structures, and (3) a unified, structure-aware pretraining framework that boosts performance across a variety of code-related tasks, particularly in code transpilation.

In addition, other than our specialized AST-aware masking approach, \model introduces no architecture changes or additional heads, and our pretraining objective remains the same as Vanilla T5. This compatibility enables seamless integration of our model as a drop-in replacement for any T5 variant.

In our experiments, \model consistently outperforms baselines in code generation, transpilation, and understanding tasks. Through controlled experiments, we empirically demonstrate that these advancements are attributed to our AST-aware pretraining techniques. Notably, \model not only outperforms similar-sized models like CodeT5 and CodeT5+ across various benchmarks but also remains competitive with, or occasionally even exceeds, the performance of much larger models using the HumanEval~\citep{codex} and the MBPP~\citep{mbpp} benchmarks. Furthermore, the inherent AST-awareness of \model offers unique advantages in structure-sensitive tasks, such as code-to-code transpilation and Clone Detection, highlighting its effectiveness at capturing the structural nuances of code.

\section{Related Work}\label{sec:related_work}



\paragraph{Language Models for Code.} Language models (LMs) extended their use from NLP to code understanding and generation. Encoder-only models generally excel in code understanding when finetuned with classifiers~\citep{codebert}, while decoder-only models are optimized for code generation through their autoregressive nature~\citep{codex,incoder,codegen}. However, these models can falter outside their primary domains of expertise or require increased resources for comparable outcomes. Our work focuses on encoder-decoder models, aiming to efficiently balance performance in both understanding and generation tasks without excessive computational demands.



\paragraph{Efforts Toward Unified Models.} Extending NLP models like BART~\citep{bart} and T5~\citep{t5}, several studies have developed encoder-decoder architectures, such as PLBART~\citep{plbart} and CodeT5~\citep{codet5}, to perform well in diverse code-related tasks. Although these models show broader utility, they struggle with generating coherent, executable code in complex scenarios like HumanEval~\citep{codex}. CodeT5+~\citep{codet5p} seeks to address this limitation through an intricate multi-task pretraining strategy across five objectives.
In contrast, our proposed model, AST-T5, uses a novel AST-Aware pretraining paradigm to become a unified model capable of generating fluent code and maintaining superior performance in code understanding tasks. Moreover, \model is more streamlined, because it only uses a single pretraining objective.


\paragraph{Leveraging Code Structure in Pretraining.}Code differs from natural language in two key aspects: its executability and strict structural syntax. Previous research leveraged execution traces for improving model performance~\citep{exec_1,exec_2,exec_3}, but this approach faces scalability challenges when applied to large, web-crawled code datasets used in pretraining. Regarding code’s structured nature, various studies have integrated syntactic elements into neural network models. \citet{struct_1}, \citet{struct_3} and \citet{struct_5} add AST-Aware attention mechanisms in their models, while \citet{struct_2} and \citet{struct_4} focus on modeling AST node expansion operations rather than traditional code tokens. In parallel, \citet{graphcodebert} and \citet{graph_1} explore DFG-Aware attention mechanisms and Graph Neural Networks (GNNs), to interpret code based on its Data Flow Graph (DFG). StructCoder~\citep{structcoder} enriches the code input by appending AST and DFG as additional features. These methods, however, necessitate parsing or static analysis for downstream tasks, which is less feasible for incomplete or incorrect code scenarios like bug fixing.

Our work, \model, aligns with methods that utilize code structure only in pretraining, like DOBF~\citep{dobf} and CodeT5~\citep{codet5}, which obfuscate inputs to force the model to grasp abstract structures. Our approach uniquely diverges by using AST-driven segmentation and masking in T5 span corruption during pretraining. This novel approach offers a more refined pretraining signal compared to structure-agnostic T5, equipping our model to proficiently encode and generate semantically coherent code structures.

\section{Method}\label{sec:method}

In this section, we present \model, a novel pretraining framework for code-based language models that harnesses the power of Abstract Syntax Trees (ASTs). First, \model parses code into ASTs to enable a deeper understanding of code structure. Leveraging this structure, we introduce AST-Aware Segmentation, an algorithm designed to address Transformer token limits while retaining the semantic coherence of the code. Second, we introduce AST-Aware Span Corruption, a masking technique that pretrains \model to reconstruct code structures ranging from individual tokens to entire function bodies, enhancing both its flexibility and structure-awareness.

\subsection{Parsing Code Into ASTs}

Unlike traditional language models on code that handle code as simple sequences of subword tokens, \model leverages the Abstract Syntax Tree (AST) of code to gain semantic insights. For parsing purposes, we assume the provided code is syntactically valid---a reasonable assumption for tasks like code transpilation and understanding. Instead of the often computationally-intensive or infeasible methods of Control-Flow Analysis (CFA) or code execution~\citep{graphcodebert,structcoder}, our method only requires the code to be parsable. We use Tree-sitter, a multi-language parser, to construct the ASTs, where each subtree represents a consecutive span of subword tokens, and every leaf node represents an individual token.

\subsection{AST-Aware Segmentation}~\label{sec:ast_aware_segmentation}


In this subsection, we describe our AST-Aware Segmentation method, which splits lengthy code files into chunks in a structure-perserving manner.

\begin{figure*}[ht]
\vskip 0.2in
\begin{center}
\centerline{\includegraphics[]{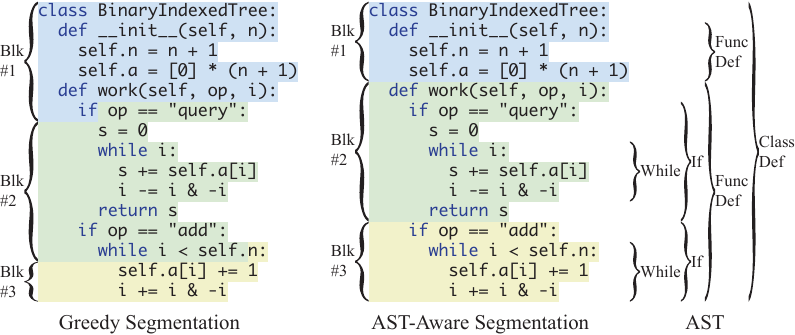}}
\caption{Comparison between Greedy Segmentation and AST-Aware Segmentation: For a 112-token code example with \texttt{max\_len} set at 48, Greedy Segmentation places the first 48 tokens in Block 1, the next 48 tokens in Block 2, and the remaining in Block 3, disrupting the structural integrity of the code. In contrast, AST-Aware Segmentation uses a dynamic programming algorithm to smartly partition the code, aligning with boundaries of member functions or major function branches, thereby preserving the code's structure. The accompanying AST, with some levels pruned for clarity, corroborates that these segmentations indeed coincide with key subtree demarcations. 
}\label{fig:ast_aware_segmentation}
\end{center}
\vskip -0.2in
\end{figure*}

{\bf Segmentation in language model pretraining} is a critical yet often overlooked aspect. Transformer LMs impose token limits on input sequences, making segmentation essential for fitting these inputs within the \texttt{max\_len} constraint. A naive approach is Greedy Segmentation, where each chunk, except the last, contains exactly \texttt{max\_len} tokens \Cref{fig:ast_aware_segmentation} (Left). This strategy has been widely adopted in previous works, such as CodeT5~\citep{codet5}.

Research in NLP by \citet{roberta} underscores that segmentation respecting sentence and document boundaries outperforms the greedy strategy. Given programming language's inherently structured nature, which is arguably more complex than natural language, a more sophisticated segmentation approach is even more important. However, this area remains largely unexplored.

{\bf AST-Aware Segmentation} is our novel approach designed to preserve the AST structure of code during segmentation. Unlike Greedy Segmentation, which can indiscriminately fragment AST structures, our method strategically minimizes such disruptions. As illustrated in the example in \Cref{fig:ast_aware_segmentation}, Greedy Segmentation leads to nine instances of AST breaks---between Block 1 and Block 2, it breaks \texttt{If}, \texttt{FuncDef}, and \texttt{ClassDef}; between Block 2 and Block 3, it breaks \texttt{Attr}, \texttt{BinaryExpr}, \texttt{While}, \texttt{If}, \texttt{FuncDef}, and \texttt{ClassDef}. In contrast, our AST-Aware approach results in only three breaks: between Block 1 and Block 2, it breaks \texttt{ClassDef}, and between Block 2 and Block 3, it breaks \texttt{FuncDef} and \texttt{ClassDef}.

To identify optimal partition boundaries, we developed the following dynamic programming (DP)-based algorithm:

\begin{enumerate}[leftmargin=*,itemsep=2pt,topsep=2pt,parsep=2pt,partopsep=2pt]

\item We construct an array \texttt{cost}, where \texttt{cost[i]} denotes the number of AST-structure breaks that would occur if partitioning happened right after token \(i\). This array is populated by traversing the AST and incrementing \texttt{cost[l..r - 1]} by 1 for each span \([l, r]\) associated with an AST subtree.

\item We define a 2-D array \texttt{dp}, where \texttt{dp[k, i]} represents the the minimum total number of AST-structure breaks when \(k\) partitions are made for the first \(i\) tokens, ending the last partition right after the \(i\)-th token. The state transition equation is:
 \begin{align} \texttt{dp}[k, i] = \texttt{cost}[i] + \min_{i - \texttt{max\_len} \le j < i}\texttt{dp}[k - 1, j] \end{align}

\item While the naive DP algorithm has a quadratic time complexity \(O(n^2)\) relative to the code file length \(n\), it can be optimized to \(O(n^2/\texttt{max\_len})\) by employing a monotonic queue for sliding-window minimum calculations. This allows for efficient computation across most code files. The pseudocode of the optimized dynamic programming algorithm is shown in \Cref{alg:dynamic_programming_opt}. See \Cref{app:segmentation} for details about complexity calculations.

\item The algorithm outputs the partition associated with \texttt{dp[k\_min, n]}, where \(\texttt{k\_min} = \arg\min_k(\texttt{dp}[k, n])\), as the most optimal partition.
\end{enumerate}

\begin{algorithm}[t]
\caption{Dynamic Programming in AST-Aware Segmentation}
\label{alg:dynamic_programming_opt}
\vskip 0.05in
\begin{minted}[linenos,xleftmargin=20pt,fontsize=\fontsize{9}{10}\selectfont]{python}
# n: the length of the code file
#    (number of tokens)
# m: the max number of segments;
#    approximately n / max_len
for k in range(1, m + 1):
  q = Queue()  # double ended queue
  for i in range(1, n + 1):
    while (q.nonempty() and
           q.left() < i - max_len):
      # pop indices before i - max_len
      q.pop_left()
    while (q.nonempty() and
           dp[k-1, q.right()] > dp[k-1, i-1]):
      # maintain monotonicity of values
      q.pop_right()
    q.push_right(i - 1)  # push i - 1
    best_j = q.left()
    # guaranteed to have the smallest value
    prev[k, i] = best_j
    dp[k, i] = cost[i] + dp[k - 1, best_j]
\end{minted}
\vskip -0.1in
\end{algorithm}

In comparing AST-Aware Segmentation with Greedy Segmentation---using the example in \Cref{fig:ast_aware_segmentation}---we find that the former presents more coherent code segments to the model during pretraining. Conversely, the latter introduces noisy partial expressions near partition boundaries. Consequently, AST-Aware Segmentation not only optimizes the pretraining process but also reduces the mismatch between pretraining and downstream tasks, which often involve complete function definitions as inputs.

\subsection{Pretraining with Span Corruption}\label{sec:method_pretraining_with_span_corruption}

\model's pretraining is based on \textit{span corruption}, a well-established method for pretraining transformer encoder-decoder models~\citep{t5}. In this approach, 15\% of the input tokens are randomly masked and replaced by unique ``sentinel'' tokens, distinct within each example. Each unique sentinel token is associated with a specific ID and added to the model's vocabulary.

During pretraining, the encoder processes the corrupted input sequence. The decoder's objective is to reconstruct the dropped-out tokens based on the encoder's output representations. Specifically, the target sequence consists of the masked spans of tokens, demarcated by their corresponding sentinel tokens. This framework effectively trains the model to recover the original text from a corrupted input. \Cref{fig:ast_aware_span_corruption} (Left) illustrates an example of the input-output pair for span corruption.

\subsection{AST-Aware Subtree Corruption}


\begin{algorithm}[t]
\caption{Subtree Selection in AST-Aware Subtree Corruption}
\label{alg:subtree_selection}
\vskip 0.05in
\begin{minted}[linenos,xleftmargin=20pt,fontsize=\fontsize{9}{10}\selectfont]{python}
def mask_subtree(t: ASTNode, m: int):
  """mask m tokens in subtree t"""
  ordered_children = []
  m_remaining = m
  # distribute m tokens among children of t
  for child in t.children:
    # theta: a hyperparameter to control
    #        masking granularity
    if child.size > theta:
      # same mask ratio as the current subtree
      m_child = m * (child.size / t.size)
      mask_subtree(child, m_child)  # recurse
      m_remaining -= m_child
    else:
      ordered_children.append(child)
  weighted_shuffle(ordered_children)
  # greedy allocation of remaining mask quota
  for child in ordered_children:
    m_child = min(m_remaining, child.size)
    mask_subtree(child, m_child)
    m_remaining -= m_child
\end{minted}
\vskip -0.1in
\end{algorithm}

\model augments the traditional span corruption paradigm by incorporating AST-awareness. Rather than arbitrarily masking consecutive token spans, \model masks code spans corresponding to AST subtrees, ranging from individual expressions to entire function bodies.

\paragraph{Subtree Masking.} We use a recursive algorithm, outlined in \Cref{alg:subtree_selection}, to traverse the AST and select subtrees for masking. The algorithm aims to fulfill two goals:

\begin{enumerate}
    \item Introduce sufficient randomness across training epochs to enhance generalization.
    \item Control the masking granularity via a tunable hyperparameter \( \theta \) (named \texttt{theta} in \Cref{alg:subtree_selection}, Line 9).
\end{enumerate}

The ``mask quota'' \( m \) denotes the number of tokens to be masked in a subtree rooted at node \( t \). The size of a subtree corresponds to the number of tokens it encompasses, derived from the cumulative sizes of its children. For larger subtrees that exceed the size threshold \( \theta \), masking is applied recursively (Lines 9-13). Meanwhile, smaller subtrees undergo a weighted shuffle, and the quota \( m \) is then apportioned among \( t \)'s children in a greedy fashion according to the shuffled order (Lines 17-21). The weights for shuffling are determined by a heuristic function on the size of each child, such that masking probabilities are distributed uniformly across leaf nodes. To create a subtree mask for an AST rooted at \(t\) with a mask ratio \(r\) (e.g., 15\% or 25\%), one can use \(\texttt{mask\_subtree}(t, \lfloor |t| \cdot r \rfloor )\).

The parameter \( \theta \) controls the granularity of masking. For example, with \( \theta = 5 \), the algorithm has a high probability to mask individual tokens and short expressions. As \( \theta \) increases to 20, the algorithm is more likely to mask larger constructs such as statements. When \( \theta = 100 \), the probability increases for masking structures like loops or entire function bodies. To foster diverse training scenarios, \( \theta \) is randomly sampled within a predefined range (e.g., 5 to 100) for each training example. This allows the pretraining framework to inherently accommodate tasks as varied as single-token completion to full function body generation from a given signature.

The subtree masking strategy is the primary distinction between our AST-Aware Subtree Corruption and the Vanilla T5 Span Corruption, as illustrated in \Cref{fig:ast_aware_span_corruption}. While conventional T5 variants mask random token spans, with an average span length of 3~\citep{t5} and neglecting code structures, our method targets the masking of AST subtrees, potentially encompassing up to 100 tokens. This equips AST-T5 for generation of various code structures coherently.

\paragraph{Pretraining Objective.} Except for the strategy used to select masked tokens and the segmentation strategy described in \Cref{sec:ast_aware_segmentation}
, our approach adheres to the workflow described in \Cref{sec:method_pretraining_with_span_corruption}. Once subtrees are selected for masking and replaced with sentinel tokens, the encoder processes this modified input. Subsequently, the decoder is tasked with reconstructing the original tokens within the masked subtrees. A side-by-side comparison between our approach and the Vanilla Span Corruption in T5 is presented in \Cref{fig:ast_aware_span_corruption}.


\section{Experimental Setup}\label{sec:experimental_setup}

\paragraph{Model Architecture.}
\model has an architecture similar to T5$_\textsc{Base}$~\citep{t5}, comprising a 12-layer encoder and a 12-layer decoder, where each layer has 768 dimensions and 12 attention heads. In total, the model has 277M parameters.

\paragraph{Pretraining.}
\model is pretrained on a subset of The Stack Dedup corpus~\citep{thestack}, a near-deduplicated version of The Stack---a 3.1TB collection of permissively licensed source code from GitHub cutoff at April 2022, spanning 358 programming languages. For our experiments, \model's training involves Python, Java, C, C++, C\#, Markdown, and reStructuredText subsets, comprising a 588GB dataset with 93M code and natural language files.

Each file is first parsed into its AST using the Tree-Sitter multi-language parser, and then tokenized with byte-level Byte-Pair Encoding (BPE) using a byte-level BPE token vocabulary.
Following AST-Aware Segmentation, these files are partitioned into chunks of 1,024 tokens. Our model is pretrained using the AST-Aware Subtree Corruption objective for 524 billion tokens (1,024 tokens per sequence, 1,024 sequences per batch, and 500k steps). For each training example, we apply AST-Aware Subtree Corruption of it is code, or apply Vanilla T5 Span Corruption of it is natural language. For code, the threshold, \( \theta \), is uniformly sampled from 5 to 100. For text, the length of each masked span is uniformly sampled from 1 to 10. Pretraining uses PyTorch, Fairseq\footnote{\url{https://github.com/facebookresearch/fairseq}} and FlashAttention~\citep{flashattention} and is conducted on 8 nodes, each with 8x NVIDIA A100 40GB GPUs. Further pretraining hyperparameters are detailed in \Cref{app:pretrain_hyperparams}.

\paragraph{Evaluation.}
We evaluate \model across three types of tasks: text-to-code generation, code-to-code transpilation, and code understanding (classification). Our evaluation encompasses tasks from the CodeXGLUE meta-benchmark~\citep{codexglue} and also includes HumanEval~\citep{codex} and MBPP~\citep{mbpp}. Specifically, for text-to-code generation, we assess performance using HumanEval, MBPP, and Concode~\citep{concode}; for transpilation, we use CodeXGLUE Java-C\# and Bugs2Fix~\citep{bugs2fix} for evaluation; and for understanding, we use BigCloneBench~\citep{bigclonebench} and the Defect Detection task proposed by \citet{devign}. Detailed metrics and statistics of these datasets are provided in \Cref{tab:benchmark}.

\begin{table}[t]
\caption{Overview of our evaluation benchmarks about test set size, task type, and evaluation metric for each task. ``Generation'' tasks involve mapping natural language to code, ``Transpilation'' tasks involve translating code from one programming language to another, and ``Understanding'' tasks involve classifying code into categorical labels. For MBPP, we follow \citet{codegen} and evaluate our model on the entire ``sanitized'' subset without few-shot prompts. For evaluation metrics, ``Pass@1'' indicates code execution on unit-tests provided in the benchmark using a single generated code per example, with reported pass rates. ``EM'' (Exact Match) evaluates textual equivalence without execution by comparing two canonicalized code pieces. ``Acc'' means accuracy in classification tasks. We omit ``BLEU scores'' because high BLEU values ($>$ 50) can still correspond to unexecutable or significantly flawed code~\citep{codexglue}, which is not useful in real-world applications. We also discuss evaluation results using the CodeBLEU~\citep{codebleu} metric in \Cref{app:codebleu}.}
\label{tab:benchmark}
\vskip 0.15in
\begin{center}\scalebox{0.95}{
\begin{tabular}{lrll}
\toprule
& \textbf{Size} & \textbf{Type} & \textbf{Metric}  \\
\midrule
HumanEval & 164 & Generation & Pass@1 \\
MBPP & 427 & Generation & Pass@1 \\
Concode & 2,000 & Generation & EM \\
Bugs2Fix & 12,379 & Transpilation & EM \\
Java-C\# & 1,000 & Transpilation & EM \\
BigCloneBench & 415,416 & Understanding & F1 \\
Defect Detect & 27,318 & Understanding & Acc \\
\bottomrule
\end{tabular}}
\end{center}
\vskip -0.1in
\end{table}

We finetune \model on the training datasets of all downstream tasks, adhering to the methodology by \citet{t5}. For the HumanEval task, which lacks its own training dataset, we use CodeSearchNet~\citep{codesearchnet}, aligning with the approach of \citet{codet5p}. The prompt templates for finetuning are constructed using the PromptSource framework~\citep{promptsource}. The finetuning takes 50k steps, with the peak learning rate set at 10\% of the pretraining learning rate. All other hyperparameters from pretraining are retained without further adjustments, and we train only one finetuned model. During inference, rank classification is employed for code understanding tasks and beam search is used for generative tasks, following \citet{t0}. For CodeXGLUE, we evaluate our model on the test set using five prompt templates for each task and report the average performance; for HumanEval and MBPP, we evaluate the top-1 generated output from beam search.

\paragraph{Baselines.}
We first benchmark \model against our own T5 baselines to ensure a controlled comparison. All models share identical Transformer architectures, pretraining data, and computational settings, differing only in the use of AST-Aware Segmentation and Subtree Corruption techniques by \model. This setup directly evaluates the efficacy of our proposed methods.

We further benchmark \model against other language models for code-related tasks. These include decoder-only models such as the GPT variants~\citep{gpt3,codex,gptj,gptneo}, PaLM~\citep{palm}, InCoder~\citep{incoder}, and LLaMa~\citep{llama}. We also compare with encoder-decoder models, including PLBART~\citep{plbart}, CodeT5~\citep{codet5}, StructCoder~\citep{structcoder}, and CodeT5+~\citep{codet5p}. Notably, CodeT5$_\textsc{Base}$ and CodeT5+ (220M) closely resemble our model in terms of architecture and size, but \model distinguishes itself with its AST-Aware pretraining techniques.

\section{Evaluation Results}\label{sec:evaluation_results}

\begin{table*}[t]
\caption{Performance comparison of various pretraining configurations for downstream tasks. Each row represents a sequential modification applied to the model in the previous row. Metrics include ``Pass@1'' rate for HumanEval, ``Exact Match'' rate for CONCODE, Bugs2Fix (for ``Small'' and ``Medium'' code lengths splits), and Java-C\# transpilation (both Java-to-C\# and C\#-to-Java). F1 score is used for Clone Detection, and Accuracy for Defect Detection, consistent with prior studies.}
\label{tab:analyses}
\vskip 0.15in
\begin{center}
\scalebox{0.95}{
\begin{tabular}{lrrrrrrr}
\toprule

                          & \multicolumn{2}{c}{\textbf{Generation}} & \multicolumn{2}{c}{\textbf{Transpilation}} & \multicolumn{2}{c}{\textbf{Understanding}} &     \\  \cmidrule(lr){2-3} \cmidrule(lr){4-5} \cmidrule(lr){6-7}
\textbf{Pretraining Config}          & \textbf{HumanEval}       & \textbf{Concode}      & \textbf{Bugs2Fix}           & \textbf{Java-C\#}          & \textbf{Clone}            & \textbf{Defect}           & \textbf{Avg} \\ \midrule
T5                        &  5.2 & 18.3 & 21.2/13.8 & 65.5/68.4 & 96.9 & 64.1 & 44.2    \\
+ AST. Segmentation       &  7.2 & 20.2 & 22.5/15.1 & 66.3/69.3 & 98.3 & 65.9 & 45.7    \\
+ AST. Subtree Corrupt    &  9.6 & 22.1 & 23.3/\textbf{16.5} & 67.3/72.2 & \textbf{98.6} & \textbf{66.0} &   47.0  \\
+ Mask 25\% (\model)      & 14.0 & \textbf{22.9} & \textbf{23.8}/16.1 & \textbf{68.9}/\textbf{72.3} & \textbf{98.6} & 65.8 & \textbf{47.9}   
\\ \midrule
+ Mask 50\%  & \textbf{14.3} & 22.0 & 21.9/15.0 & 66.5/70.1 & 97.1 & 64.2 & 46.4
\\ \bottomrule
\end{tabular}}
\end{center}
\vskip -0.1in
\end{table*}

In this section, we evaluate AST-T5 across multiple benchmarks. First, we analyze the contributions of each component within our AST-aware pretraining framework through controlled experiments. Next, we benchmark AST-T5 against existing models in prior work.

\subsection{Pretraining Procedure Analysis}

In this subsection, we analyze the key components that contribute to the pretraining of \model models. Holding the model architecture, pretraining datasets, and computational environment constant, we sequentially add one component at a time to a T5 baseline trained on code, culminating in our finalized \model model. \Cref{tab:analyses} presents the experimental results. These results show that:

\paragraph{AST-Aware Segmentation enhances code language models.} A comparison between the first two rows of \Cref{tab:analyses} shows that the model trained with AST-Aware Segmentation consistently outperforms the T5 baseline that uses Greedy Segmentation across all tasks. The advantage stems from the fact that AST-Aware Segmentation produces less fragmented and thus less noisy training inputs during pretraining. Given that most downstream tasks present coherent code structures, such as entire function definitions, the consistency upheld by AST-Aware pretraining aligns better with these structures, leading to improved generalization.

\paragraph{AST-Aware Span Corruption further boosts generation performance.} A comparison between the second and third rows of \Cref{tab:analyses} reveals an improvement when shifting from Vanilla T5 Span Corruption to our AST-Aware Subtree Corruption. This performance gain is especially notable in generation and transpilation tasks. Such enhancements stem from the ability of AST-Aware Subtree Corruption to guide the model in generating code with better coherence and structural integrity.

\paragraph{Increasing masking ratio improves generation performance.} The typical span corruption mask ratio in T5 is set at 15\%. Increasing this ratio could potentially enhance the model's generation capabilities, albeit potentially at the expense of understanding tasks. Essentially, a mask ratio of 100\% would emulate a GPT-like, decoder-only Transformer. However, in our experiments (last two rows of \Cref{tab:analyses}), we observed that raising the mask ratio from 15\% to 25\% significantly improved generation capabilities without noticeably compromising performance in understanding tasks. Further analysis shows that increasing the masking ratio to 50\% yields only a marginal improvement on HumanEval (from 14.0 to 14.3), while adversely impacting transpilation and understanding tasks. Thus, we settled on a 25\% mask ratio for our \model model.

\subsection{Main Results}

\begin{table*}[t]
\caption{Results of \model on downstream tasks compared with reported results of established language models. Evaluation metrics align with those in Table 1. Our focus is primarily on models with similar sizes as \model, specifically the ``Base'' models (100M to 300M parameters), while comparisons against larger models are depicted in Figure 3. Some models are either encoder-only or decoder-only and are thus not suited for certain tasks. These results are labeled with ``\textcolor{gray}{N/A}'' in this table because they are not available in the literature.}
\label{tab:main}
\vskip 0.15in
\begin{center}
\scalebox{0.94}{
\begin{tabular}{lrrrrrrr}
\toprule
& \multicolumn{2}{c}{\textbf{Generation}} & \multicolumn{2}{c}{\textbf{Transpilation}} & \multicolumn{2}{c}{\textbf{Understanding}}    \\  \cmidrule(lr){2-3} \cmidrule(lr){4-5} \cmidrule(lr){6-7}
\textbf{Model}  &  \textbf{HumanEval}       & \textbf{Concode}      & \textbf{Bugs2Fix}           & \textbf{Java-C\#}          & \textbf{Clone}            & \textbf{Defect}   \\ \midrule
CodeBERT &   \textcolor{gray}{N/A} & \textcolor{gray}{N/A} & 16.4 / 5.2 & 59.0/58.8 & 96.5 & 62.1 \\
GraphCodeBERT &  \textcolor{gray}{N/A} &  \textcolor{gray}{N/A} & 17.3 / 9.1 & 59.4/58.8 &  97.1 &  \textcolor{gray}{N/A} \\
PLBART &  \textcolor{gray}{N/A} & 18.8 &  19.2 / 9.0 & 64.6/65.0 & 97.2 & 63.2 \\
CodeT5 &  \textcolor{gray}{N/A} & 22.3 &  21.6/14.0& 65.9/66.9 &  97.2 & 65.8 \\
CodeT5+$_\textsc{Base}$ &  12.0 &  \textcolor{gray}{N/A} &  \textcolor{gray}{N/A}  &  \textcolor{gray}{N/A} & 95.2 & \textbf{66.1} \\
StructCoder &   \textcolor{gray}{N/A} & 22.4 &  \textcolor{gray}{N/A} & 66.9/68.7 &  \textcolor{gray}{N/A} &  \textcolor{gray}{N/A} \\
AST-T5 (Ours) &  \textbf{14.0} & \textbf{22.9} & \textbf{23.8}/\textbf{16.1} & \textbf{68.9}/\textbf{72.3} & \textbf{98.6} & 65.8 \\ \bottomrule
\end{tabular}}
\end{center}
\vskip -0.1in
\end{table*}

\begin{figure*}[t]
\vskip 0.2in
    \begin{center}
    \begin{subfigure}[b]{0.49\textwidth}
        \centering
        \includegraphics[width=2.9in]{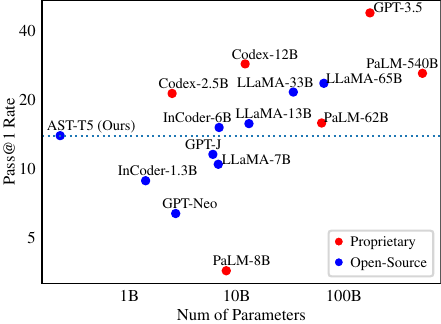}
        \caption{HumanEval}
        \label{fig:humaneval}
    \end{subfigure}
    \hfill
    \begin{subfigure}[b]{0.49\textwidth}
        \centering
        \includegraphics[width=2.9in]{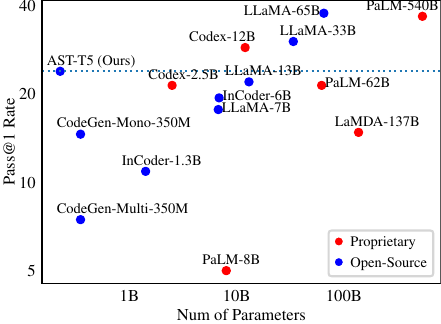}
        \caption{MBPP}
        \label{fig:mbpp}
    \end{subfigure}
    \caption{Visualizations of \model's performance on HumanEval and MBPP compared to other models compared to models exceeding 300M parameters. Each point on each scatter plot represents a model. The x-axis shows the parameter count in log-scale, while the y-axis shows the Pass@1 rate on HumanEval or MBPP in log-scale. Model open-source status is color-coded: \textcolor{blue}{\textbf{blue}} for open-source and \textcolor{red}{\textbf{red}} for proprietary.}
    \end{center}
    \vskip -0.2in
\end{figure*}

\Cref{tab:main} shows \model's performance on downstream tasks compared with previously published results of similarly sized models, specifically those within the ``Base'' scale (100M to 300M parameters). \Cref{fig:humaneval} and \Cref{fig:mbpp} extends this comparison, comparing \model with larger models using the HumanEval benchmark and the MBPP benchmark, respectively. Additional results on EvalPlus are shown in \Cref{app:evalplus}. These results show that:

\paragraph{\model excels as a unified and parameter-efficient LM for various code-related tasks.} While comparable in size, \model consistently outperforms similar-sized models such as CodeT5~\citep{codet5} and CodeT5+~\citep{codet5p} in code generation, transpilation, and understanding. Notably, while CodeT5 and CodeT5+ are models at the Base scale, they were evaluated across different tasks. Our model, \model, outperforms the best results of these two models across multiple benchmarks at the same time. Moreover, \Cref{fig:humaneval} highlights \model's competitiveness against significantly larger models like GPT-J~\citep{gptj} and LLaMa-7B~\citep{llama} on the HumanEval benchmark, underscoring our model's parameter efficiency. Similarly, \Cref{fig:mbpp} demonstrates \model's advantages over LLaMa-7B and Codex-2.5B~\citep{codex} on the MBPP benchmark, showing the effectiveness of \model.

\paragraph{\model exhibits unique strengths in transpilation through AST-awareness.} \Cref{tab:main} highlights \model's superior performance in code-to-code transpilation tasks, showcasing gains a substantial gain of 2 to 5 points on Bugs2Fix and Java-C\# transpilation. In transpilation, while surface-level code can exhibit significant variability, the intrinsic AST structures of the source and target often maintain a notable similarity. The capability of \model to exploit this structural similarity is crucial to its effectiveness. The benefits of being structure-aware are further exemplified by \model's leading results in Clone Detection, where it surpasses CodeT5 by 3 points, because AST comparisons yield more precise insights than direct code comparisons.

\section{Conclusion and Future Work}\label{sec:conclusion}

In this work, we present \model, a novel pretraining paradigm that harnesses the power of Abstract Syntax Trees (ASTs) to boost the performance of code-centric language models. Using two structure-aware techniques, \model not only outperforms models of comparable size but also competes favorably against some larger counterparts. The simplicity of \model lies in its singular pretraining objective and its adaptability as a drop-in replacement for any encoder-decoder LM, highlighting its potential for real-world deployments. Moving forward, we aim to explore the scalability of \model by training larger models on more expansive datasets.

\section*{Acknowledgements}

We thank the reviewers and metareviewers at ICML for their constructive feedback and support. This work is supported in part by gift from Meta, the U.S. National Science Foundation through grants IIS-1955488, IIS-2027575, ARO W911NF2110339, ONR N00014-21-1-2724, and DOE awards DE-SC0016260, DE-SC0021982. The AI training platform and computation resources supporting this work were provided by High-Flyer AI Fundamental Research Co. Ltd, Hangzhou, China. We thank Sida Wang, Chris Cummins, Volker Seeker, and Hugh Leather for their valuable discussions.

\section*{Impact Statement}

In this paper, we introduce \model, a language model aimed at automated generation, transpilation, and understanding of code. The advancement of LLMs in code generation raises concerns about automated code production's security, privacy, and potential misuse. There is a risk that improved code generation capabilities could be exploited for malicious purposes, such as automating the creation of software vulnerabilities or facilitating the development of harmful software. Our research emphasizes the importance of responsible AI development and use, advocating for continuous monitoring, ethical guidelines, and safeguards to mitigate these risks.

\bibliography{references}

\begin{thebibliography}{47}
\providecommand{\natexlab}[1]{#1}
\providecommand{\url}[1]{\texttt{#1}}
\expandafter\ifx\csname urlstyle\endcsname\relax
  \providecommand{\doi}[1]{doi: #1}\else
  \providecommand{\doi}{doi: \begingroup \urlstyle{rm}\Url}\fi

\bibitem[Ahmad et~al.(2021)Ahmad, Chakraborty, Ray, and Chang]{plbart}
Ahmad, W.~U., Chakraborty, S., Ray, B., and Chang, K.-W.
\newblock Unified pre-training for program understanding and generation.
\newblock Apr 2021.
\newblock \doi{10.48550/arXiv.2103.06333}.
\newblock URL \url{http://arxiv.org/abs/2103.06333}.
\newblock arXiv:2103.06333 [cs].

\bibitem[Allamanis et~al.(2017)Allamanis, Brockschmidt, and Khademi]{graph_1}
Allamanis, M., Brockschmidt, M., and Khademi, M.
\newblock Learning to represent programs with graphs.
\newblock Nov 2017.
\newblock URL \url{https://arxiv.org/abs/1711.00740}.
\newblock arXiv:1711.00740 [cs].

\bibitem[Alon et~al.(2020)Alon, Sadaka, Levy, and Yahav]{struct_2}
Alon, U., Sadaka, R., Levy, O., and Yahav, E.
\newblock Structural language models of code.
\newblock July 2020.
\newblock \doi{10.48550/arXiv.1910.00577}.
\newblock URL \url{http://arxiv.org/abs/1910.00577}.
\newblock arXiv:1910.00577 [cs, stat].

\bibitem[Athiwaratkun et~al.(2023)Athiwaratkun, Gouda, Wang, Li, Tian, Tan, Ahmad, Wang, Sun, Shang, Gonugondla, Ding, Kumar, Fulton, Farahani, Jain, Giaquinto, Qian, Ramanathan, Nallapati, Ray, Bhatia, Sengupta, Roth, and Xiang]{mbxp}
Athiwaratkun, B., Gouda, S.~K., Wang, Z., Li, X., Tian, Y., Tan, M., Ahmad, W.~U., Wang, S., Sun, Q., Shang, M., Gonugondla, S.~K., Ding, H., Kumar, V., Fulton, N., Farahani, A., Jain, S., Giaquinto, R., Qian, H., Ramanathan, M.~K., Nallapati, R., Ray, B., Bhatia, P., Sengupta, S., Roth, D., and Xiang, B.
\newblock Multi-lingual evaluation of code generation models.
\newblock \penalty0 (arXiv:2210.14868), March 2023.
\newblock URL \url{http://arxiv.org/abs/2210.14868}.
\newblock arXiv:2210.14868 [cs].

\bibitem[Austin et~al.(2021)Austin, Odena, Nye, Bosma, Michalewski, Dohan, Jiang, Cai, Terry, Le, and Sutton]{mbpp}
Austin, J., Odena, A., Nye, M., Bosma, M., Michalewski, H., Dohan, D., Jiang, E., Cai, C., Terry, M., Le, Q., and Sutton, C.
\newblock Program synthesis with large language models.
\newblock Aug 2021.
\newblock \doi{10.48550/arXiv.2108.07732}.
\newblock URL \url{http://arxiv.org/abs/2108.07732}.
\newblock arXiv:2108.07732 [cs].

\bibitem[Bach et~al.(2022)Bach, Sanh, Yong, Webson, Raffel, Nayak, Sharma, Kim, Bari, Fevry, Alyafeai, Dey, Santilli, Sun, Ben-David, Xu, Chhablani, Wang, Fries, Al-shaibani, Sharma, Thakker, Almubarak, Tang, Radev, Jiang, and Rush]{promptsource}
Bach, S.~H., Sanh, V., Yong, Z.-X., Webson, A., Raffel, C., Nayak, N.~V., Sharma, A., Kim, T., Bari, M.~S., Fevry, T., Alyafeai, Z., Dey, M., Santilli, A., Sun, Z., Ben-David, S., Xu, C., Chhablani, G., Wang, H., Fries, J.~A., Al-shaibani, M.~S., Sharma, S., Thakker, U., Almubarak, K., Tang, X., Radev, D., Jiang, M. T.-J., and Rush, A.~M.
\newblock {P}rompt{S}ource: An integrated development environment and repository for natural language prompts.
\newblock March 2022.
\newblock \doi{10.48550/arXiv.2202.01279}.
\newblock URL \url{http://arxiv.org/abs/2202.01279}.
\newblock arXiv:2202.01279 [cs].

\bibitem[BigScience(2021)]{bloom}
BigScience.
\newblock {B}igscience {L}anguage {O}pen-science {O}pen-access {M}ultilingual ({BLOOM}), May 2021.
\newblock URL \url{https://huggingface.co/bigscience/bloom}.

\bibitem[Black et~al.(2021)Black, Gao, Wang, Leahy, and Biderman]{gptneo}
Black, S., Gao, L., Wang, P., Leahy, C., and Biderman, S.
\newblock {GPT-Neo: Large Scale Autoregressive Language Modeling with Mesh-Tensorflow}, March 2021.
\newblock URL \url{https://doi.org/10.5281/zenodo.5297715}.

\bibitem[Brown et~al.(2020)Brown, Mann, Ryder, Subbiah, Kaplan, Dhariwal, Neelakantan, Shyam, Sastry, Askell, Agarwal, Herbert-Voss, Krueger, Henighan, Child, Ramesh, Ziegler, Wu, Winter, Hesse, Chen, Sigler, Litwin, Gray, Chess, Clark, Berner, McCandlish, Radford, Sutskever, and Amodei]{gpt3}
Brown, T.~B., Mann, B., Ryder, N., Subbiah, M., Kaplan, J., Dhariwal, P., Neelakantan, A., Shyam, P., Sastry, G., Askell, A., Agarwal, S., Herbert-Voss, A., Krueger, G., Henighan, T., Child, R., Ramesh, A., Ziegler, D.~M., Wu, J., Winter, C., Hesse, C., Chen, M., Sigler, E., Litwin, M., Gray, S., Chess, B., Clark, J., Berner, C., McCandlish, S., Radford, A., Sutskever, I., and Amodei, D.
\newblock Language models are few-shot learners.
\newblock Jul 2020.
\newblock \doi{10.48550/arXiv.2005.14165}.
\newblock URL \url{http://arxiv.org/abs/2005.14165}.
\newblock arXiv:2005.14165 [cs].

\bibitem[Chen et~al.(2021{\natexlab{a}})Chen, Tworek, Jun, Yuan, Pinto, Kaplan, Edwards, Burda, Joseph, Brockman, Ray, Puri, Krueger, Petrov, Khlaaf, Sastry, Mishkin, Chan, Gray, Ryder, Pavlov, Power, Kaiser, Bavarian, Winter, Tillet, Such, Cummings, Plappert, Chantzis, Barnes, Herbert-Voss, Guss, Nichol, Paino, Tezak, Tang, Babuschkin, Balaji, Jain, Saunders, Hesse, Carr, Leike, Achiam, Misra, Morikawa, Radford, Knight, Brundage, Murati, Mayer, Welinder, McGrew, Amodei, McCandlish, Sutskever, and Zaremba]{codex}
Chen, M., Tworek, J., Jun, H., Yuan, Q., Pinto, H. P. d.~O., Kaplan, J., Edwards, H., Burda, Y., Joseph, N., Brockman, G., Ray, A., Puri, R., Krueger, G., Petrov, M., Khlaaf, H., Sastry, G., Mishkin, P., Chan, B., Gray, S., Ryder, N., Pavlov, M., Power, A., Kaiser, L., Bavarian, M., Winter, C., Tillet, P., Such, F.~P., Cummings, D., Plappert, M., Chantzis, F., Barnes, E., Herbert-Voss, A., Guss, W.~H., Nichol, A., Paino, A., Tezak, N., Tang, J., Babuschkin, I., Balaji, S., Jain, S., Saunders, W., Hesse, C., Carr, A.~N., Leike, J., Achiam, J., Misra, V., Morikawa, E., Radford, A., Knight, M., Brundage, M., Murati, M., Mayer, K., Welinder, P., McGrew, B., Amodei, D., McCandlish, S., Sutskever, I., and Zaremba, W.
\newblock Evaluating large language models trained on code.
\newblock Jul 2021{\natexlab{a}}.
\newblock \doi{10.48550/arXiv.2107.03374}.
\newblock URL \url{http://arxiv.org/abs/2107.03374}.
\newblock arXiv:2107.03374 [cs].

\bibitem[Chen et~al.(2018)Chen, Liu, and Song]{exec_1}
Chen, X., Liu, C., and Song, D.
\newblock Execution-guided neural program synthesis.
\newblock Sep 2018.
\newblock URL \url{https://openreview.net/forum?id=H1gfOiAqYm}.

\bibitem[Chen et~al.(2021{\natexlab{b}})Chen, Song, and Tian]{exec_2}
Chen, X., Song, D., and Tian, Y.
\newblock Latent execution for neural program synthesis.
\newblock Jun 2021{\natexlab{b}}.
\newblock URL \url{https://arxiv.org/abs/2107.00101}.
\newblock arXiv:2107.00101 [cs].

\bibitem[Chowdhery et~al.(2022)Chowdhery, Narang, Devlin, Bosma, Mishra, Roberts, Barham, Chung, Sutton, Gehrmann, Schuh, Shi, Tsvyashchenko, Maynez, Rao, Barnes, Tay, Shazeer, Prabhakaran, Reif, Du, Hutchinson, Pope, Bradbury, Austin, Isard, Gur-Ari, Yin, Duke, Levskaya, Ghemawat, Dev, Michalewski, Garcia, Misra, Robinson, Fedus, Zhou, Ippolito, Luan, Lim, Zoph, Spiridonov, Sepassi, Dohan, Agrawal, Omernick, Dai, Pillai, Pellat, Lewkowycz, Moreira, Child, Polozov, Lee, Zhou, Wang, Saeta, Diaz, Firat, Catasta, Wei, Meier-Hellstern, Eck, Dean, Petrov, and Fiedel]{palm}
Chowdhery, A., Narang, S., Devlin, J., Bosma, M., Mishra, G., Roberts, A., Barham, P., Chung, H.~W., Sutton, C., Gehrmann, S., Schuh, P., Shi, K., Tsvyashchenko, S., Maynez, J., Rao, A., Barnes, P., Tay, Y., Shazeer, N., Prabhakaran, V., Reif, E., Du, N., Hutchinson, B., Pope, R., Bradbury, J., Austin, J., Isard, M., Gur-Ari, G., Yin, P., Duke, T., Levskaya, A., Ghemawat, S., Dev, S., Michalewski, H., Garcia, X., Misra, V., Robinson, K., Fedus, L., Zhou, D., Ippolito, D., Luan, D., Lim, H., Zoph, B., Spiridonov, A., Sepassi, R., Dohan, D., Agrawal, S., Omernick, M., Dai, A.~M., Pillai, T.~S., Pellat, M., Lewkowycz, A., Moreira, E., Child, R., Polozov, O., Lee, K., Zhou, Z., Wang, X., Saeta, B., Diaz, M., Firat, O., Catasta, M., Wei, J., Meier-Hellstern, K., Eck, D., Dean, J., Petrov, S., and Fiedel, N.
\newblock {P}a{LM}: Scaling language modeling with pathways.
\newblock Oct 2022.
\newblock \doi{10.48550/arXiv.2204.02311}.
\newblock URL \url{http://arxiv.org/abs/2204.02311}.
\newblock arXiv:2204.02311 [cs].

\bibitem[Dao et~al.(2022)Dao, Fu, Ermon, Rudra, and Ré]{flashattention}
Dao, T., Fu, D.~Y., Ermon, S., Rudra, A., and Ré, C.
\newblock {F}lash{A}ttention: Fast and memory-efficient exact attention with {IO}-awareness.
\newblock June 2022.
\newblock \doi{10.48550/arXiv.2205.14135}.
\newblock URL \url{http://arxiv.org/abs/2205.14135}.
\newblock arXiv:2205.14135 [cs].

\bibitem[Feng et~al.(2020)Feng, Guo, Tang, Duan, Feng, Gong, Shou, Qin, Liu, Jiang, and Zhou]{codebert}
Feng, Z., Guo, D., Tang, D., Duan, N., Feng, X., Gong, M., Shou, L., Qin, B., Liu, T., Jiang, D., and Zhou, M.
\newblock {C}ode{BERT}: A pre-trained model for programming and natural languages.
\newblock Sep 2020.
\newblock \doi{10.48550/arXiv.2002.08155}.
\newblock URL \url{http://arxiv.org/abs/2002.08155}.
\newblock arXiv:2002.08155 [cs].

\bibitem[Fried et~al.(2023)Fried, Aghajanyan, Lin, Wang, Wallace, Shi, Zhong, Yih, Zettlemoyer, and Lewis]{incoder}
Fried, D., Aghajanyan, A., Lin, J., Wang, S., Wallace, E., Shi, F., Zhong, R., Yih, W.-t., Zettlemoyer, L., and Lewis, M.
\newblock {I}n{C}oder: A generative model for code infilling and synthesis.
\newblock Apr 2023.
\newblock \doi{10.48550/arXiv.2204.05999}.
\newblock URL \url{http://arxiv.org/abs/2204.05999}.
\newblock arXiv:2204.05999 [cs].

\bibitem[Guo et~al.(2021)Guo, Ren, Lu, Feng, Tang, Liu, Zhou, Duan, Svyatkovskiy, Fu, Tufano, Deng, Clement, Drain, Sundaresan, Yin, Jiang, and Zhou]{graphcodebert}
Guo, D., Ren, S., Lu, S., Feng, Z., Tang, D., Liu, S., Zhou, L., Duan, N., Svyatkovskiy, A., Fu, S., Tufano, M., Deng, S.~K., Clement, C., Drain, D., Sundaresan, N., Yin, J., Jiang, D., and Zhou, M.
\newblock {G}raph{C}ode{BERT}: Pre-training code representations with data flow.
\newblock Sep 2021.
\newblock \doi{10.48550/arXiv.2009.08366}.
\newblock URL \url{http://arxiv.org/abs/2009.08366}.
\newblock arXiv:2009.08366 [cs].

\bibitem[Husain et~al.(2020)Husain, Wu, Gazit, Allamanis, and Brockschmidt]{codesearchnet}
Husain, H., Wu, H.-H., Gazit, T., Allamanis, M., and Brockschmidt, M.
\newblock {C}ode{S}earch{N}et challenge: Evaluating the state of semantic code search.
\newblock Jun 2020.
\newblock \doi{10.48550/arXiv.1909.09436}.
\newblock URL \url{http://arxiv.org/abs/1909.09436}.
\newblock arXiv:1909.09436 [cs, stat].

\bibitem[Iyer et~al.(2018)Iyer, Konstas, Cheung, and Zettlemoyer]{concode}
Iyer, S., Konstas, I., Cheung, A., and Zettlemoyer, L.
\newblock Mapping language to code in programmatic context.
\newblock Aug 2018.
\newblock \doi{10.48550/arXiv.1808.09588}.
\newblock URL \url{http://arxiv.org/abs/1808.09588}.
\newblock arXiv:1808.09588 [cs].

\bibitem[Kim et~al.(2021)Kim, Zhao, Tian, and Chandra]{struct_3}
Kim, S., Zhao, J., Tian, Y., and Chandra, S.
\newblock Code prediction by feeding trees to transformers.
\newblock March 2021.
\newblock \doi{10.48550/arXiv.2003.13848}.
\newblock URL \url{http://arxiv.org/abs/2003.13848}.
\newblock arXiv:2003.13848 [cs].

\bibitem[Kocetkov et~al.(2022)Kocetkov, Li, Allal, Li, Mou, Ferrandis, Jernite, Mitchell, Hughes, Wolf, Bahdanau, von Werra, and de~Vries]{thestack}
Kocetkov, D., Li, R., Allal, L.~B., Li, J., Mou, C., Ferrandis, C.~M., Jernite, Y., Mitchell, M., Hughes, S., Wolf, T., Bahdanau, D., von Werra, L., and de~Vries, H.
\newblock The {S}tack: 3 {TB} of permissively licensed source code.
\newblock \penalty0 (arXiv:2211.15533), November 2022.
\newblock \doi{10.48550/arXiv.2211.15533}.
\newblock URL \url{http://arxiv.org/abs/2211.15533}.
\newblock arXiv:2211.15533 [cs].

\bibitem[Lachaux et~al.(2020)Lachaux, Roziere, Chanussot, and Lample]{transcoder}
Lachaux, M.-A., Roziere, B., Chanussot, L., and Lample, G.
\newblock Unsupervised translation of programming languages.
\newblock Sep 2020.
\newblock \doi{10.48550/arXiv.2006.03511}.
\newblock URL \url{http://arxiv.org/abs/2006.03511}.
\newblock arXiv:2006.03511 [cs].

\bibitem[Lewis et~al.(2019)Lewis, Liu, Goyal, Ghazvininejad, Mohamed, Levy, Stoyanov, and Zettlemoyer]{bart}
Lewis, M., Liu, Y., Goyal, N., Ghazvininejad, M., Mohamed, A., Levy, O., Stoyanov, V., and Zettlemoyer, L.
\newblock {BART}: Denoising sequence-to-sequence pre-training for natural language generation, translation, and comprehension.
\newblock Oct 2019.
\newblock \doi{10.48550/arXiv.1910.13461}.
\newblock URL \url{http://arxiv.org/abs/1910.13461}.
\newblock arXiv:1910.13461 [cs, stat].

\bibitem[Li et~al.(2018)Li, Wang, Lyu, and King]{struct_1}
Li, J., Wang, Y., Lyu, M.~R., and King, I.
\newblock Code completion with neural attention and pointer networks.
\newblock In \emph{Proceedings of the Twenty-Seventh International Joint Conference on Artificial Intelligence}, pp.\  4159–4165, July 2018.
\newblock \doi{10.24963/ijcai.2018/578}.
\newblock URL \url{http://arxiv.org/abs/1711.09573}.
\newblock arXiv:1711.09573 [cs].

\bibitem[Liu et~al.(2023)Liu, Xia, Wang, and Zhang]{evalplus}
Liu, J., Xia, C.~S., Wang, Y., and Zhang, L.
\newblock Is your code generated by chat{GPT} really correct? rigorous evaluation of large language models for code generation.
\newblock In \emph{Thirty-seventh Conference on Neural Information Processing Systems}, 2023.
\newblock URL \url{https://openreview.net/forum?id=1qvx610Cu7}.

\bibitem[Liu et~al.(2019)Liu, Ott, Goyal, Du, Joshi, Chen, Levy, Lewis, Zettlemoyer, and Stoyanov]{roberta}
Liu, Y., Ott, M., Goyal, N., Du, J., Joshi, M., Chen, D., Levy, O., Lewis, M., Zettlemoyer, L., and Stoyanov, V.
\newblock {R}o{BERT}a: A robustly optimized {BERT} pretraining approach.
\newblock Jul 2019.
\newblock \doi{10.48550/arXiv.1907.11692}.
\newblock URL \url{http://arxiv.org/abs/1907.11692}.
\newblock arXiv:1907.11692 [cs].

\bibitem[Lu et~al.(2021)Lu, Guo, Ren, Huang, Svyatkovskiy, Blanco, Clement, Drain, Jiang, Tang, Li, Zhou, Shou, Zhou, Tufano, Gong, Zhou, Duan, Sundaresan, Deng, Fu, and Liu]{codexglue}
Lu, S., Guo, D., Ren, S., Huang, J., Svyatkovskiy, A., Blanco, A., Clement, C., Drain, D., Jiang, D., Tang, D., Li, G., Zhou, L., Shou, L., Zhou, L., Tufano, M., Gong, M., Zhou, M., Duan, N., Sundaresan, N., Deng, S.~K., Fu, S., and Liu, S.
\newblock {C}ode{XGLUE}: A machine learning benchmark dataset for code understanding and generation.
\newblock Mar 2021.
\newblock \doi{10.48550/arXiv.2102.04664}.
\newblock URL \url{http://arxiv.org/abs/2102.04664}.
\newblock arXiv:2102.04664 [cs].

\bibitem[Nijkamp et~al.(2023{\natexlab{a}})Nijkamp, Hayashi, Xiong, Savarese, and Zhou]{codegen2}
Nijkamp, E., Hayashi, H., Xiong, C., Savarese, S., and Zhou, Y.
\newblock {C}ode{G}en2: Lessons for training {LLM}s on programming and natural languages.
\newblock \penalty0 (arXiv:2305.02309), July 2023{\natexlab{a}}.
\newblock \doi{10.48550/arXiv.2305.02309}.
\newblock URL \url{http://arxiv.org/abs/2305.02309}.
\newblock arXiv:2305.02309 [cs].

\bibitem[Nijkamp et~al.(2023{\natexlab{b}})Nijkamp, Pang, Hayashi, Tu, Wang, Zhou, Savarese, and Xiong]{codegen}
Nijkamp, E., Pang, B., Hayashi, H., Tu, L., Wang, H., Zhou, Y., Savarese, S., and Xiong, C.
\newblock {C}ode{G}en: An open large language model for code with multi-turn program synthesis.
\newblock Feb 2023{\natexlab{b}}.
\newblock \doi{10.48550/arXiv.2203.13474}.
\newblock URL \url{http://arxiv.org/abs/2203.13474}.
\newblock arXiv:2203.13474 [cs].

\bibitem[Ouyang et~al.(2022)Ouyang, Wu, Jiang, Almeida, Wainwright, Mishkin, Zhang, Agarwal, Slama, Ray, Schulman, Hilton, Kelton, Miller, Simens, Askell, Welinder, Christiano, Leike, and Lowe]{instructgpt}
Ouyang, L., Wu, J., Jiang, X., Almeida, D., Wainwright, C.~L., Mishkin, P., Zhang, C., Agarwal, S., Slama, K., Ray, A., Schulman, J., Hilton, J., Kelton, F., Miller, L., Simens, M., Askell, A., Welinder, P., Christiano, P., Leike, J., and Lowe, R.
\newblock Training language models to follow instructions with human feedback.
\newblock Mar 2022.
\newblock \doi{10.48550/arXiv.2203.02155}.
\newblock URL \url{http://arxiv.org/abs/2203.02155}.
\newblock arXiv:2203.02155 [cs].

\bibitem[Rabinovich et~al.(2017)Rabinovich, Stern, and Klein]{struct_4}
Rabinovich, M., Stern, M., and Klein, D.
\newblock Abstract syntax networks for code generation and semantic parsing.
\newblock April 2017.
\newblock \doi{10.48550/arXiv.1704.07535}.
\newblock URL \url{http://arxiv.org/abs/1704.07535}.
\newblock arXiv:1704.07535 [cs, stat].

\bibitem[Raffel et~al.(2020)Raffel, Shazeer, Roberts, Lee, Narang, Matena, Zhou, Li, and Liu]{t5}
Raffel, C., Shazeer, N., Roberts, A., Lee, K., Narang, S., Matena, M., Zhou, Y., Li, W., and Liu, P.~J.
\newblock Exploring the limits of transfer learning with a unified text-to-text transformer.
\newblock Jul 2020.
\newblock \doi{10.48550/arXiv.1910.10683}.
\newblock URL \url{http://arxiv.org/abs/1910.10683}.
\newblock arXiv:1910.10683 [cs, stat].

\bibitem[Ren et~al.(2020)Ren, Guo, Lu, Zhou, Liu, Tang, Sundaresan, Zhou, Blanco, and Ma]{codebleu}
Ren, S., Guo, D., Lu, S., Zhou, L., Liu, S., Tang, D., Sundaresan, N., Zhou, M., Blanco, A., and Ma, S.
\newblock {C}ode{BLEU}: a method for automatic evaluation of code synthesis.
\newblock \penalty0 (arXiv:2009.10297), September 2020.
\newblock \doi{10.48550/arXiv.2009.10297}.
\newblock URL \url{http://arxiv.org/abs/2009.10297}.
\newblock arXiv:2009.10297 [cs].

\bibitem[Roziere et~al.(2021)Roziere, Lachaux, Szafraniec, and Lample]{dobf}
Roziere, B., Lachaux, M.-A., Szafraniec, M., and Lample, G.
\newblock {DOBF}: A deobfuscation pre-training objective for programming languages.
\newblock Oct 2021.
\newblock \doi{10.48550/arXiv.2102.07492}.
\newblock URL \url{http://arxiv.org/abs/2102.07492}.
\newblock arXiv:2102.07492 [cs].

\bibitem[Rozière et~al.(2023)Rozière, Gehring, Gloeckle, Sootla, Gat, Tan, Adi, Liu, Remez, Rapin, Kozhevnikov, Evtimov, Bitton, Bhatt, Ferrer, Grattafiori, Xiong, Défossez, Copet, Azhar, Touvron, Martin, Usunier, Scialom, and Synnaeve]{codellama}
Rozière, B., Gehring, J., Gloeckle, F., Sootla, S., Gat, I., Tan, X.~E., Adi, Y., Liu, J., Remez, T., Rapin, J., Kozhevnikov, A., Evtimov, I., Bitton, J., Bhatt, M., Ferrer, C.~C., Grattafiori, A., Xiong, W., Défossez, A., Copet, J., Azhar, F., Touvron, H., Martin, L., Usunier, N., Scialom, T., and Synnaeve, G.
\newblock Code llama: Open foundation models for code.
\newblock Aug 2023.
\newblock \doi{10.48550/arXiv.2308.12950}.
\newblock URL \url{http://arxiv.org/abs/2308.12950}.
\newblock arXiv:2308.12950 [cs].

\bibitem[Sanh et~al.(2021)Sanh, Webson, Raffel, Bach, Sutawika, Alyafeai, Chaffin, Stiegler, Scao, Raja, Dey, Bari, Xu, Thakker, Sharma, Szczechla, Kim, Chhablani, Nayak, Datta, Chang, Jiang, Wang, Manica, Shen, Yong, Pandey, Bawden, Wang, Neeraj, Rozen, Sharma, Santilli, Fevry, Fries, Teehan, Bers, Biderman, Gao, Wolf, and Rush]{t0}
Sanh, V., Webson, A., Raffel, C., Bach, S.~H., Sutawika, L., Alyafeai, Z., Chaffin, A., Stiegler, A., Scao, T.~L., Raja, A., Dey, M., Bari, M.~S., Xu, C., Thakker, U., Sharma, S.~S., Szczechla, E., Kim, T., Chhablani, G., Nayak, N., Datta, D., Chang, J., Jiang, M. T.-J., Wang, H., Manica, M., Shen, S., Yong, Z.~X., Pandey, H., Bawden, R., Wang, T., Neeraj, T., Rozen, J., Sharma, A., Santilli, A., Fevry, T., Fries, J.~A., Teehan, R., Bers, T., Biderman, S., Gao, L., Wolf, T., and Rush, A.~M.
\newblock Multitask prompted training enables zero-shot task generalization.
\newblock \emph{arXiv.org}, Oct 2021.
\newblock URL \url{https://arxiv.org/abs/2110.08207v3}.

\bibitem[Shojaee et~al.(2023)Shojaee, Jain, Tipirneni, and Reddy]{exec_3}
Shojaee, P., Jain, A., Tipirneni, S., and Reddy, C.~K.
\newblock Execution-based code generation using deep reinforcement learning.
\newblock Jan 2023.
\newblock URL \url{https://arxiv.org/abs/2301.13816}.
\newblock arXiv:2301.13816 [cs].

\bibitem[Svajlenko et~al.(2014)Svajlenko, Islam, Keivanloo, Roy, and Mia]{bigclonebench}
Svajlenko, J., Islam, J.~F., Keivanloo, I., Roy, C.~K., and Mia, M.~M.
\newblock Towards a big data curated benchmark of inter-project code clones.
\newblock In \emph{2014 IEEE International Conference on Software Maintenance and Evolution}, pp.\  476–480, Sep 2014.
\newblock \doi{10.1109/ICSME.2014.77}.

\bibitem[Tipirneni et~al.(2023)Tipirneni, Zhu, and Reddy]{structcoder}
Tipirneni, S., Zhu, M., and Reddy, C.~K.
\newblock {S}truct{C}oder: Structure-aware transformer for code generation.
\newblock May 2023.
\newblock \doi{10.48550/arXiv.2206.05239}.
\newblock URL \url{http://arxiv.org/abs/2206.05239}.
\newblock arXiv:2206.05239 [cs].

\bibitem[Touvron et~al.(2023)Touvron, Lavril, Izacard, Martinet, Lachaux, Lacroix, Rozière, Goyal, Hambro, Azhar, Rodriguez, Joulin, Grave, and Lample]{llama}
Touvron, H., Lavril, T., Izacard, G., Martinet, X., Lachaux, M.-A., Lacroix, T., Rozière, B., Goyal, N., Hambro, E., Azhar, F., Rodriguez, A., Joulin, A., Grave, E., and Lample, G.
\newblock {LL}a{MA}: Open and efficient foundation language models.
\newblock Feb 2023.
\newblock \doi{10.48550/arXiv.2302.13971}.
\newblock URL \url{http://arxiv.org/abs/2302.13971}.
\newblock arXiv:2302.13971 [cs].

\bibitem[Tufano et~al.(2019)Tufano, Watson, Bavota, Di~Penta, White, and Poshyvanyk]{bugs2fix}
Tufano, M., Watson, C., Bavota, G., Di~Penta, M., White, M., and Poshyvanyk, D.
\newblock An empirical study on learning bug-fixing patches in the wild via neural machine translation.
\newblock May 2019.
\newblock \doi{10.48550/arXiv.1812.08693}.
\newblock URL \url{http://arxiv.org/abs/1812.08693}.
\newblock arXiv:1812.08693 [cs].

\bibitem[Wang \& Komatsuzaki(2021)Wang and Komatsuzaki]{gptj}
Wang, B. and Komatsuzaki, A.
\newblock {GPT}-{J}-6{B}: 6{B} {JAX}-based {T}ransformer, Jun 2021.
\newblock URL \url{https://arankomatsuzaki.wordpress.com/2021/06/04/gpt-j/}.

\bibitem[Wang et~al.(2021)Wang, Wang, Joty, and Hoi]{codet5}
Wang, Y., Wang, W., Joty, S., and Hoi, S. C.~H.
\newblock {C}ode{T}5: Identifier-aware unified pre-trained encoder-decoder models for code understanding and generation.
\newblock Sep 2021.
\newblock \doi{10.48550/arXiv.2109.00859}.
\newblock URL \url{http://arxiv.org/abs/2109.00859}.
\newblock arXiv:2109.00859 [cs].

\bibitem[Wang et~al.(2023)Wang, Le, Gotmare, Bui, Li, and Hoi]{codet5p}
Wang, Y., Le, H., Gotmare, A.~D., Bui, N. D.~Q., Li, J., and Hoi, S. C.~H.
\newblock {C}ode{T}5+: Open code large language models for code understanding and generation.
\newblock May 2023.
\newblock \doi{10.48550/arXiv.2305.07922}.
\newblock URL \url{http://arxiv.org/abs/2305.07922}.
\newblock arXiv:2305.07922 [cs].

\bibitem[Zhang et~al.(2022)Zhang, Roller, Goyal, Artetxe, Chen, Chen, Dewan, Diab, Li, Lin, Mihaylov, Ott, Shleifer, Shuster, Simig, Koura, Sridhar, Wang, and Zettlemoyer]{opt}
Zhang, S., Roller, S., Goyal, N., Artetxe, M., Chen, M., Chen, S., Dewan, C., Diab, M., Li, X., Lin, X.~V., Mihaylov, T., Ott, M., Shleifer, S., Shuster, K., Simig, D., Koura, P.~S., Sridhar, A., Wang, T., and Zettlemoyer, L.
\newblock {OPT}: Open pre-trained transformer language models.
\newblock \penalty0 (arXiv:2205.01068), June 2022.
\newblock \doi{10.48550/arXiv.2205.01068}.
\newblock URL \url{http://arxiv.org/abs/2205.01068}.
\newblock arXiv:2205.01068 [cs].

\bibitem[Zhou et~al.(2019)Zhou, Liu, Siow, Du, and Liu]{devign}
Zhou, Y., Liu, S., Siow, J., Du, X., and Liu, Y.
\newblock {D}evign: Effective vulnerability identification by learning comprehensive program semantics via graph neural networks.
\newblock Sep 2019.
\newblock \doi{10.48550/arXiv.1909.03496}.
\newblock URL \url{http://arxiv.org/abs/1909.03496}.
\newblock arXiv:1909.03496 [cs, stat].

\bibitem[Zügner et~al.(2021)Zügner, Kirschstein, Catasta, Leskovec, and Günnemann]{struct_5}
Zügner, D., Kirschstein, T., Catasta, M., Leskovec, J., and Günnemann, S.
\newblock Language-agnostic representation learning of source code from structure and context.
\newblock March 2021.
\newblock \doi{10.48550/arXiv.2103.11318}.
\newblock URL \url{http://arxiv.org/abs/2103.11318}.
\newblock arXiv:2103.11318 [cs].

\end{thebibliography}
\bibliographystyle{icml2024}

\newpage
\appendix
\onecolumn

\section{Appendix}

\subsection{Limitations}

\model is specifically designed to enhance code generation performance by exclusively masking code within AST subtrees during pretraining. While this specialized approach is advantageous for code generation tasks, it may result in suboptimal performance in natural language generation. Acknowledging this limitation, future versions of \model could investigate strategies such as masking docstrings and comments to broaden its applicability. This would potentially improve performance across various tasks, including code summarization.

\subsection{More about AST-Aware Segmentation}\label{app:segmentation}

In \Cref{sec:ast_aware_segmentation}, we use a dynamic programming algorithm to calculate the segmentation that results in the least number of AST structure breaks. A naive implementation of the DP algorithm is shown in \Cref{alg:dynamic_programming}.

\begin{algorithm}
\caption{Dynamic Programming in AST-Aware Segmentation (Before Optimization)}
\label{alg:dynamic_programming}
\begin{minted}[linenos,xleftmargin=20pt]{python}
for k in range(1, m + 1):
    for i in range(1, n + 1):
        best_j = i - max_len
        for j in range(i - max_len + 1, i):
            if dp[k - 1, j] < dp[k - 1, best_j]:
                best_j = j
        prev[k, i] = best_j
        dp[k, i] = cost[i] + min_value
\end{minted}
\end{algorithm}

Denote the length of the code file (in tokens) by $n$. In the algorithm, $m$ denotes the maximum number of chunks that the file can be split into, which is approximately $n / \mathrm{max\_len}$. So this implementation has time complexity $O(m n \cdot \mathrm{max\_len}) = O(n^2)$, which is not feasible for longer code files. To optimize this algorithm, we use a monotonic queue to compute the sliding-window minimum, as described in \Cref{alg:dynamic_programming_opt}.

Each element is only pushed into and popped out of the monotonic queue once, so the time complexity of the optimized algorithm is $O(nm) = O(n^2 / \mathrm{max\_len})$, making the algorithm \(\sim\) 1000x faster when $\mathrm{max\_len} = 1024$. This allows the algorithm to segment each code file with 100k tokens in milliseconds.

\subsection{Pretraining Hyperparameters}\label{app:pretrain_hyperparams}

\Cref{tab:appendix_hparam_pretrain} shows the pretraining hyperparameters for our proposed \model model.

\begin{table}[h]
\centering
\begin{tabular}{lr}
\toprule
Encoder Layers & 12  \\
Decoder Layers & 12  \\
Hidden Dimension & 768  \\
Peak Learning Rate & 2e-4  \\
Batch Size & 1,024  \\
Warm-Up Steps & 10,000  \\
Total Steps & 500,000 \\
Sequence Length & 1,024 \\
Mask Ratio & 25\% \\
Min Subtree Corruption Threshold \(\theta\) & 5 \\
Max Subtree Corruption Threshold \(\theta\) & 100 \\
Relative Position Encoding Buckets & 32  \\
Relative Position Encoding Max Distance & 128  \\
Adam $\epsilon$ & 1e-6  \\
Adam ($\beta_1$, $\beta_2$) & (0.9, 0.98)  \\
Clip Norm &  2.0  \\
Dropout & 0.1   \\
Weight Decay & 0.01  \\
\bottomrule
\end{tabular}
\caption{Pretraining hyperparameters for our \model model.}
\label{tab:appendix_hparam_pretrain}
\end{table}





\subsection{Evaluation Results on EvalPlus}\label{app:evalplus}

We extend our evaluation to include EvalPlus~\citep{evalplus}, a more rigorous benchmark that enhances the original HumanEval and MBPP datasets with a substantial number of additional test cases. EvalPlus is designed to provide a more accurate evaluation of the correctness of programs produced by LLMs.

For our tests on HumanEval+ and MBPP+, we use the same hyperparameters used in our evaluations of HumanEval and MBPP. It is important to note that the hyperparameter configurations used in our study are not directly comparable to those used for the models listed on the EvalPlus leaderboard\footnote{\url{https://evalplus.github.io/leaderboard.html}}. Our results are compared against established models including GPT-Neo, GPT-J, InCoder, and CodeGen-2~\citep{codegen2}.

\begin{table}[t]
\caption{Performance of \model on HumanEval+ and MBPP+ benchmarks, compared with reported numbers of language models listed on the EvalPlus leaderboard. The evaluation metric used is Pass@1.}
\label{tab:evalplus}
\begin{center}
\begin{tabular}{lrrr}
\toprule
              & \textbf{\#Params} &  \textbf{HumanEval+} & \textbf{MBPP+} \\ \midrule
GPT-Neo & 2.7B & 6.7 & 7.9 \\
GPT-J & 6B & 11.0 & 12.2 \\
InCoder-1.3B & 1.3B & 11.0 & 12.2 \\
InCoder-6.7B & 6.7B & 12.2 & 15.9 \\
CodeGen2-1B & 1B & 9.1 & 11.0 \\
CodeGen2-3B & 3B & 12.8 & 15.9 \\
CodeGen2-7B & 7B & 17.7 & 18.3 \\
CodeGen2-16B & 16B & 16.5 & 19.5 \\
AST-T5 (Ours) & 277M & 12.8 & 19.3 \\ \bottomrule
\end{tabular}
\end{center}
\end{table}

As shown in \Cref{tab:evalplus}, our 277M-parameter \model outperforms larger models like InCoder-6.7B and CodeGen2-1B, showing the effectiveness and parameter efficiency of \model.

\subsection{Evaluation Results on Multi-Lingual Code Generation}\label{app:multieval}

\begin{table}[t]
\caption{Results of \model on multi-lingual HumanEval and MBXP compared with reported results of established language models. The evaluation metric is Pass@1.}
\label{tab:multilingual}
\begin{center}
\begin{tabular}{lrrrrr}
\toprule
              & \textbf{\#Params} & \multicolumn{2}{c}{\textbf{HumanEval}} & \multicolumn{2}{c}{\textbf{MBXP}} \\
              &          & Python                & Java                & Python       & Java      \\ \midrule
CodeGen-multi & 350M      & 7.3                   & 5.0                 & 7.5          & 8.2       \\
CodeGen-mono  & 350M      & 10.3                  & 3.1                 & \textbf{14.6}         & 1.9       \\
AST-T5 (Ours) & 277M     & 14.0                  & \textbf{10.6}                & \textbf{23.9}         & \textbf{9.8}       \\ \midrule
BLOOM         & 7.1B      & 7.9                   & 8.1                 & 7.0          & 7.8       \\
OPT           & 13B      & 0.6                   & 0.6                 & 1.4          & 1.4       \\
CodeGen-multi & 2B     & 11.0                  & 11.2                & 18.8         & 19.5      \\
CodeGen-mono  & 2B      & 20.7                  & 5.0                 & 31.7         & 16.7      \\
CodeGen-multi & 6B     & 15.2                  & 10.6                & 22.5         & 21.7      \\
CodeGen-mono  & 6B      & 19.5                  & 8.7                 & 37.2         & 19.8      \\
CodeGen-multi & 16B     & 17.1                  & 16.2                & 24.2         & 28.0      \\
CodeGen-mono  & 16B     & 22.6                  & 22.4                & 40.6         & 26.8      \\ \bottomrule
\end{tabular}
\end{center}
\end{table}

\Cref{tab:multilingual} presents a comparative analysis of our \model model on Python and Java subsets of the multi-lingual HumanEval and MBXP benchmarks \citep{mbxp}. This analysis includes models such as BLOOM~\citep{bloom}, OPT~\citep{opt}, and various configurations of CodeGen \citep{codegen}, as reported in \citet{mbxp}. Our results show \model's superior performance across all benchmarks compared to the CodeGen-multi-350M.
Furthermore, AST-T5, having 277M parameters, outperforms larger counterparts like BLOOM-7.1B and OPT-13B.

\subsection{Evaluation Results in CodeBLEU}\label{app:codebleu}

\begin{table}[t]
\caption{Results of \model on CONCODE with reported results of established language models. The evaluation metric is exact match score and CodeBLEU.}
\label{tab:codebleu}
\vskip 0.15in
\begin{center}
\begin{tabular}{lrr}
\toprule
                & \textbf{EM}   & \textbf{CodeBLEU} \\ \midrule
GPT-2           & 17.4 & 29.7     \\
CodeGPT-2       & 18.3 & 32.7     \\
CodeGPT-adapted & 20.1 & 36.0     \\
PLBART          & 18.8 & 38.5     \\
CodeT5-Small    & 21.6 & 41.4     \\
CodeT5-Base     & 22.3 & 43.2     \\
AST-T5 (Ours)   & \textbf{22.9} & \textbf{45.0}    \\ \bottomrule
\end{tabular}
\end{center}
\vskip -0.1in
\end{table}

\Cref{tab:codebleu} presents the performance of various models on the Concode dataset using the CodeBLEU metric, as reported in \citep{codet5}. CodeBLEU, specifically designed for evaluating code synthesis, computes a weighted average of three scores: textual match (BLEU), AST match, and Data Flow Graph (DFG) match. Our findings show a clear correlation between CodeBLEU and exact match scores.

\end{document}